\documentclass[]{taira}

\usepackage{graphicx}
\usepackage{epstopdf, epsfig}
\usepackage{graphicx}
\usepackage{dcolumn}
\usepackage{bm}
\usepackage{color}
\usepackage{array}
\usepackage{tabu}
\usepackage{amsmath,amssymb,stmaryrd}
\usepackage{multicol}
\usepackage{algorithm}
\usepackage[noend]{algpseudocode}
\usepackage{subfigure}
\usepackage{subfigmat}
\usepackage{pbox}
\usepackage{xfrac}
\usepackage{tabularx}
\usepackage{multirow}
\usepackage[percent]{overpic}
\usepackage{tikz}
\usepackage[nomessages]{fp}

\definecolor{blue}{rgb}{0, 0.4470, 0.7410}
\definecolor{red}{rgb}{0.8500, 0.1250, 0.0480} 
\definecolor{orange}{rgb}{0.8500, 0.3250, 0.0980} 
\definecolor{yellow}{rgb}{0.9290, 0.6940, 0.1250}
\definecolor{purple}{rgb}{0.4940, 0.1840, 0.5560}
\definecolor{green}{rgb}{0.4660, 0.6740, 0.1880}
\definecolor{ltblue}{rgb}{0.3010, 0.7450, 0.9330}
\definecolor{dkred}{rgb}{0.6350, 0.0780, 0.1840}
\definecolor{gray}{rgb}{0.22, 0.22, 0.3}
\definecolor{ltgray}{rgb}{0.8, 0.8, 0.8}
\definecolor{magenta}{rgb}{1, 0, 1}
\definecolor{oldblue}{rgb}{0, 0, 1}
\definecolor{oldred}{rgb}{1, 0, 0} 

\newcolumntype{M}[1]{>{\centering\arraybackslash}m{#1}}

\makeatletter
\def\BState{\State\hskip-\ALG@thistlm}
\makeatother

\shorttitle{Broadcast analysis of 2D turbulence}
\title{Network broadcast analysis and control of turbulent flows}

\shortauthor{C.-A. Yeh, M. Gopalakrishnan Meena \& K. Taira}
\author{Chi-An Yeh\corresp{\email{cayeh@seas.ucla.edu}}, 
Muralikrishnan Gopalakrishnan Meena\aunote{Present address: National Center for Computational Sciences, Oak Ridge National Laboratory},
and \\
Kunihiko Taira}

\affiliation{Mechanical and Aerospace Engineering, University of California, Los Angeles, CA 90095, USA}

\begin{document}

\maketitle

\begin{abstract}
We present a network-based modal analysis technique that identifies key dynamical paths along which perturbations amplify over a time-varying base flow.  This analysis is built upon the Katz centrality, which reveals the flow structures that can effectively spread perturbations over a time-evolving network of vortical interactions on the base flow.  Motivated by the resolvent form of the Katz function, we take the singular value decomposition of the resulting communicability matrix, complementing the resolvent analysis for fluid flows.  The right-singular vectors, referred to as the broadcast modes, give insights into the sensitive regions where introduced perturbations can be effectively spread and amplified over the entire fluid-flow network that evolves in time.  We apply this analysis to a two-dimensional decaying isotropic turbulence.  The broadcast mode reveals that vortex dipoles are important structures in spreading perturbations.  By perturbing the flow with the principal broadcast mode,  we demonstrate the utility of the insights gained from the present analysis to effectively modify the evolution of turbulent flows.  The current network-inspired work presents a novel use of network analysis to guide flow control efforts, in particular for time-varying base flows.
\end{abstract}



\section{Introduction}

In a sea of vortices, a dense network of vortical interactions gives rise to their complex dynamics.  For the characterization, modeling, and control of such vortical flow, the identification of flow structures influential to the overall governing dynamics is critical \citep{Hunt:CTR1988, Haller:ARFM2015, Jeong:JFM1995, Jimenez:JFM2018, Jimenez:JT2020}.  Due to the large degrees of freedom in describing complex vortical flows, tremendous efforts have been placed on utilizing modal analysis techniques to extract the key flow structures in a low-order representation \citep{Holmes:2012, Schmid:JFM2010, Rowley:IJBC2005, Theofilis:ARFM2011, Taira:AIAAJ2017, Taira:AIAAJ2020}.  Identification of these structures is important from the standpoint of energy, dynamics, control, stability, and input-output characteristics.  By revealing the influential structures, the actuation can be applied wisely to modify the overall vortical flow dynamics efficiently and effectively, providing pathways to improve operations of fluid based engineering systems.

Analogous problems of information and disease transmissions over networks are studied in the field of network science \citep{Newman:SIAMRev2003, Newman:2018, Dorogovtsev:2010, Barabasi:2016}.  Among many network science problems, the spread of diseases over a human network and the propagation of information over the internet have been extensively studied \citep{Broder:CN2000, Liljeros:Nature2001, Albert:RMP2002, Barabasi:SciA2003, Brockmann:Science2013, Firth:Nature2020}.  The identification of critical nodes in a network for these problems is important from the standpoint of control, security, and public health.  These nodes are often found using network centralities that quantify their connectivities in terms of their ability to broadcast or receive information (or disease) over the network.  Naturally, these concepts from network science can be related to actuator and sensor placement problems in flow control. The question of which node could initiate a broadcast such that information can widely spread over a social network is analogous to where to place an actuator in a flow for effective amplification of control input.

The use of network-theoretic tools for analyzing fluid flows has been rapidly emerging over the past few years, considering various types of interaction that form a network according to the applications and the physical mechanisms of interest.  The Lagrangian motion of fluid elements has been used to quantify interactions in fluid flows \citep{SerGiacomi:Chaos2015, Iacobello:JFM2019}.  Time series of fluid flow properties have been considered to establish visibility graphs and recurrence networks, extracting dynamical features of complex flows \citep{Scarsoglio:Chaos2017, Godavarthi:Chao2017}.  Network analysis has been used to study triadic interactions in turbulent flows \citep{Gurcan:PRE2017} and to model vortical interactions in Lagrangian and Eulerian settings \citep{Nair:JFM2015, Taira:JFM2016, MGM:PRE2018}.  Network-based frameworks have also been utilized to identify influential structures in complex flows using techniques such as spectral clustering \citep{Hadjighasem:PRE2016}, coherent structure coloring \citep{Schlueter-Kuck:JFM2017}, community detection \citep{Murayama:PRE2018, MGM:JFM2020}, and graph comparison \citep{Krueger:JFM2019}.  Most of these frameworks rely on a time-frozen or kinematic approach to analyze flows that evolve in time, hence they are limited in the identification of time-dependent critical regions in fluid flows.  Moreover, although these techniques identify regions for effective broadcast for modifying vortical flows, they give little guidance on the receiving characteristics of the network-based modifications.

To address these issues, this study introduces a time-evolving network framework to identify the dynamically relevant pathways of vortical interactions, with the goal of effectively modifying the flow and providing guidance on where, when, and how such modifications will affect the flow evolution.  We present an analytical approach that combines the toolsets from network analysis and modal analysis to identify influential structures in time-varying vortical flows.  Through the lens of a time-evolving network of vortical interactions, we use network centrality measures to reveal the sensitive regions where the added perturbation can be effectively amplified and modify the flow.  The centrality measure of our particular interest is the Katz centrality \citep{Katz:Psychometrika1953}, which has been regarded as a highly insightful measure specifically for identifying important nodes in time-evolving networks \citep{Grindrod:PRE2011}.  Motivated by the expression of Katz centrality, we combine it with concepts from resolvent analysis \citep{Trefethen:Science1993, Jovanovic:JFM2005, McKeonSharma:JFM2010, Yeh:JFM2019} to characterize the input-output relationship over a time-evolving turbulent flow.  We refer to this blended formulation as the {\it broadcast analysis}, which serves as a systematic approach for analyzing time-varying base flows that remain challenging to most modal analysis techniques.  

To demonstrate the strength of the present approach in analyzing time-varying fluid flows, we apply the current analysis to two-dimensional (2D) decaying isotropic turbulence \citep{Boffetta:ARFM2012}.  The 2D turbulence is chosen as a benchmark problem, since its characteristics have been studied in numbers of studies \citep{Lilly:JFM1971, McWilliams:JFM1990, Jimenez:JFM2018, Jimenez:JT2020}.  However, the control of 2D turbulence remains challenging, due to its its highly unsteady and chaotic nature with minimal coherence.  Therefore, our goal is to use the broadcast analysis to identify the important structures in 2D turbulence and their capability of modifying its complex evolution.  With a successful application of the present approach for turbulent flow modifications, our hope is to pave a way towards effective control of time-varying flow with high levels of unsteadiness.

In what follows, we present the broadcast analysis and its implications for studying time-varying networks/flows in section \ref{sec:approach}.  The current formulation is then used to analyze and modify two-dimensional isotropic turbulence as an example in sections \ref{sec:setups} and \ref{sec:results}.  At last, concluding remarks are offered in section \ref{sec:conclusion}.

\section{Broadcast analysis}
\label{sec:approach}
\subsection{Time-evolving network}

A network is defined by a set of nodes connected by edges holding weights to quantify the strengths of the connections \citep{Newman:2018, Barabasi:2016, Dorogovtsev:2010}.  The {\it adjacency matrix} $\mathsfbi{A} \in \mathbb{R}^{n \times n}$ uniquely describes a network of $n$ nodes, with each entry $\mathsfi{A}_{ij}$ being the edge weight that quantifies the influence of node $j$ on node $i$.  For a time-evolving network, the connections between the nodes, or edge weights, vary in time, yielding a time-dependent $\mathsfbi{A}(t)$.  

\begin{figure}
\vspace{0.5in}
\centering
\begin{overpic}[width=0.98\textwidth]{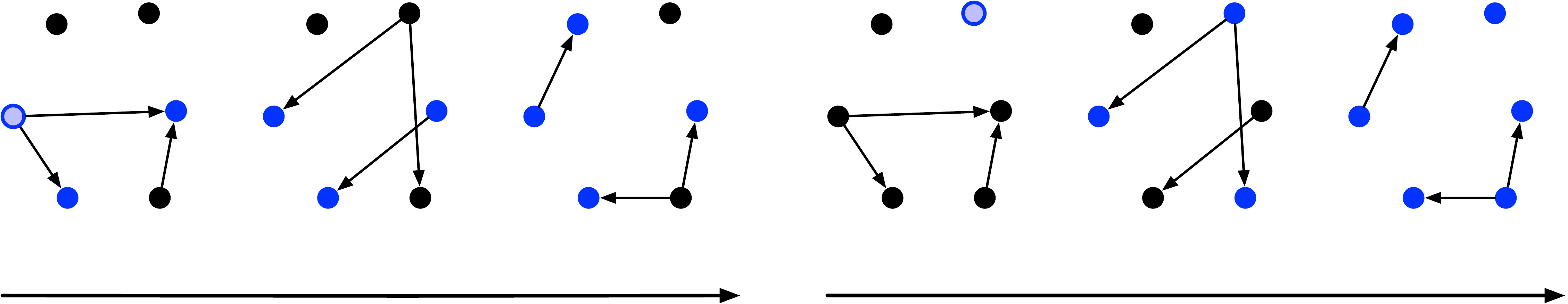}
	\put(00.0, 25.0){\small (a) Seed message at node A}
	
	\put(-1.0, 13.1){\indexsize A}
	\put( 2.5, 19.1){\indexsize B}
	\put( 9.1, 19.7){\indexsize C}
	\put(11.0, 13.3){\indexsize D}
	\put(11.0, 05.2){\indexsize E}
	\put( 2.1, 05.2){\indexsize F}
	
	\put( 4.8, 03.0){\indexsize $t = t_1$}
	\put(21.5, 03.0){\indexsize $t = t_2$}
	\put(38.2, 03.0){\indexsize $t = t_3$}
	\put(42.2, -2.0){\indexsize Time}
	
	\put(52.0, 25.0){\small (b) Seed message at node C}

	\put(51.7, 13.1){\indexsize\selectfont A}
	\put(55.2, 19.1){\indexsize\selectfont B}
	\put(61.8, 19.7){\indexsize\selectfont C}
	\put(63.7, 13.3){\indexsize\selectfont D}
	\put(63.7, 05.2){\indexsize\selectfont E}
	\put(54.8, 05.2){\indexsize\selectfont F}
	
	\put(57.5, 03.0){\indexsize $t = t_1$}
	\put(74.2, 03.0){\indexsize $t = t_2$}
	\put(90.9, 03.0){\indexsize $t = t_3$}
	\put(94.9, -2.0){\indexsize Time}
\end{overpic}
\vspace{0.15in}
\caption{\label{fig:time_evo_network_example}  For the same time-evolving network, message is seeded at A and C at $t_1$ for (a) and (b), respectively.  Message seeded at node C is more widely spread by $t_3$ than seeded at A, even though node A at $t_1$ has the most outgoing edges.}
\end{figure}
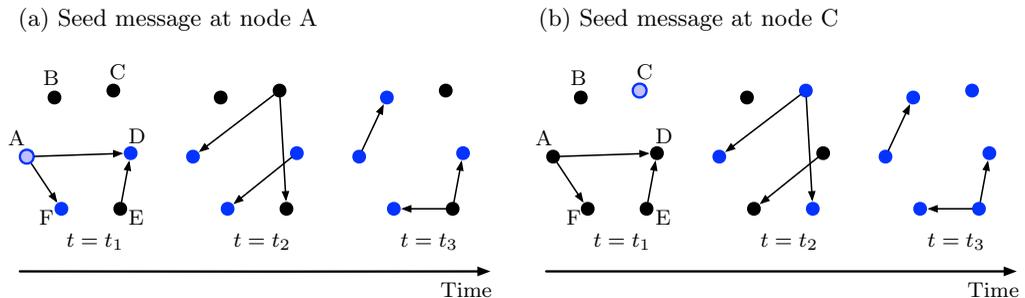

Let us consider a simple example in figure \ref{fig:time_evo_network_example} to analyze a time-evolving network and show how the influential nodes can be drastically different when we take time evolution into account.  This example models a communication network with directed edges between the nodes being one-way information transfers.  For this evolving network, we seek the node to seed a message at $t = t_1$ such that it can be received by the largest number of nodes at $t = t_3$.  The propagation of the message from nodes A and C at $t = t_1$ are shown in figure \ref{fig:time_evo_network_example} (a) and (b), respectively.  If we consider the network at $t = t_1$ as static, node A is the most influential node in spreading the message, as it possesses the most outgoing edges, or the highest {\it out degree} \citep{Newman:2018}.  The out degree serves as an effective centrality measure for static networks.  However, when considering the time-evolving network, the nodes that receive the message (marked in blue) at $t = t_3$ are in fact fewer than those when the message was initially seeded at node C, which has no connections at $t_1$.  This observation motivates the use of an alternative measure for time-evolving networks.

We also note that, since node C has no connections at $t_1$, the seeded message only stays at node C without being passed.  The option for the information to stay at a node until better connections appear at a later time is important in determining effective nodes for broadcasting information \citep{Grindrod:PRE2011}.  This option is fulfilled with the use of the Katz centrality for identifying important nodes in a time-evolving network.  

\subsection{Katz centrality and walk downweighting}

The concept of Katz centrality is based on the combinatorics of {\it walks} via which a distributed information, $\boldsymbol{f} \in \mathbb{R}^n$, can be spread over the network \citep{Katz:Psychometrika1953,Grindrod:PRE2011}.  Here, a walk is defined as a single pass of the information from one node to another if there is an edge connecting them.  Hence, the propagation of information after one walk can be represented by $\boldsymbol{q} = \mathsfbi{A}\boldsymbol{f}$, where $\boldsymbol{q} \in \mathbb{R}^n$ is the amount of information aggregated to each node after the walk.  For a fluid-flow network, the information $\boldsymbol{f}$ and $\boldsymbol{q}$ can be respectively interpreted as a perturbation to be spread and the global response over the flow field.

The extraction of Katz centrality considers the process where $\boldsymbol{f}$ is passed over the nodes by any amount of walks, namely $\mathsfbi{A}^p\boldsymbol{f}$ with $p \in \mathbb{N}$.  Along each walk, the amount of information transferred from node $j$ to $i$ is scaled by the edge weight $\mathsfi{A}_{ij}$ and a walk-downweighting parameter $\alpha$.  Accounting for all paths comprised of any amount of walks, this process of information transfer can be expressed as 
\begin{equation}
\label{eq:Katz_series}
	\boldsymbol{q} = (\boldsymbol{I} + \alpha\mathsfbi{A} + \alpha^2\mathsfbi{A}^2 + \cdots) \boldsymbol{f}, 
\end{equation}
where we note that the identity matrix represents the option for the information to stay at the same node without being passed to others.  While the value of $\alpha$ can be chosen by matching other centrality measures \citep{Aprahamian:JCN2016}, in this study we will choose its value based on the physics of the problem.  Moreover, when $\alpha$ satisfies $\alpha < 1/\rho(\mathsfbi{A})$, where $\rho(\mathsfbi{A})$ is the spectral radius of $\mathsfbi{A}$, the infinite series in (\ref{eq:Katz_series}) converges such that

\begin{equation}
\label{eq:Katz_resolvent}
	\boldsymbol{q} = \mathsfbi{K} \boldsymbol{f},
\end{equation}
where
\begin{equation}
\label{eq:Katz_func}
	\mathsfbi{K} \equiv \left(\boldsymbol{I} - \alpha\mathsfbi{A} \right)^{-1}
\end{equation}
is referred to as the Katz function.  By taking the column sum of the Katz function, the traditional Katz broadcast centrality is computed as $\boldsymbol{b}_{\text{K}} = \boldsymbol{1} (\boldsymbol{I} - \alpha\mathsfbi{A} )^{-1}$  with $\boldsymbol{1} = [1, 1, \dots, 1]$, which quantifies the ability of each node to broadcast information to all others in the network. 

We note that the Katz function $\mathsfbi{K}(\mathsfbi{A},\alpha) \equiv \left(\boldsymbol{I} - \alpha\mathsfbi{A} \right)^{-1}$ is in the resolvent form of the adjacency matrix $\mathsfbi{A}$.  It also captures the input--output process between $\boldsymbol{q}$ and $\boldsymbol{f}$ as shown in equation (\ref{eq:Katz_resolvent}).  Therefore, rather than taking the column sum to compute the broadcast centrality, we adopt the resolvent analysis formulation \citep{Trefethen:Science1993} and perform the singular value decomposition (SVD) of the Katz function as
\begin{equation}
\label{eq:Katz_SVD}
	\mathsfbi{K}(\mathsfbi{A},\alpha) = \mathsfbi{R}\boldsymbol{\Sigma}\mathsfbi{B},
\end{equation}  
where we refer to the leading right-singular vector $\boldsymbol{b}$ in $\mathsfbi{B}$ as the {\it broadcast mode}.  This broadcast mode also complements the concept of forcing mode in resolvent analysis, as it identifies the most influential nodes, or sensitive regions in the flow, that can effectively spread perturbations over the vortical network.  Furthermore, the leading left-singular vector $\boldsymbol{r}$ in $\mathsfbi{R}$ can be interpreted as the {\it receiving mode}, which accounts for the information aggregating to each node when originally distributed in the shape of $\boldsymbol{b}$.  Compared to the traditional Katz centrality, the use of SVD not only reveals the important nodes in broadcasting and receiving information, but also identifies the optimal way to distribute information over the nodes with amplification corresponding to the singular value $\sigma$ when being spread through the operation of $\mathsfbi{K}(\mathsfbi{A},\alpha)$.

\subsection{Communicability matrix and age downweighting}

In the analysis of a time-evolving network, the adjacency matrix is usually sampled at discrete times as a series of $\mathsfbi{A}_k$ instead of the time-continuous form $\mathsfbi{A}(t)$.  The network $\mathsfbi{A}_k$ sampled at $t_k$ can be considered time-frozen over the finite time increment $\Delta t_k$ until the next sample of adjacency matrix $\mathsfbi{A}_{k+1}$ becomes available at $t_{k+1} = t_k + \Delta t_k$ \citep{Grindrod:PRE2011}.  Over this finite time-increment $\Delta t_k$, a perturbation can spread over the network by taking multiple walks over the time-frozen $\mathsfbi{A}_k$, considering each walk takes $\delta t \ll \Delta t_k$, before it propagates to the next time slice where the network evolves to $\mathsfbi{A}_{k+1}$.  Thus, the propagation of perturbation over a time-evolving network can be evaluated as

\begin{equation}
\label{eq:comm_series}
	\boldsymbol{q}(t_m) = 
	\underbrace{(\boldsymbol{I} + \alpha\mathsfbi{A}_m + \alpha^2\mathsfbi{A}_m^2 + \cdots)}_{t = t_m}
	\underbrace{\left(\cdots\right)~\cdots~\left(\cdots\right)}_{t \in [t_{m-1},\cdots,t_2,t_1]}
	\underbrace{(\boldsymbol{I} + \alpha\mathsfbi{A}_0 + \alpha^2\mathsfbi{A}_0^2 + \cdots)}_{t = t_0}
	\boldsymbol{f}(t_0). 
\end{equation}
With an appropriate choice of the walk-downweighting parameter $\alpha$, we can use the Katz function for each time slice and describe the propagation of information over the time horizon of $t \in [t_0, t_m]$ as 

\begin{equation}
\label{eq:io_comm}
	\boldsymbol{q}(t_m) = \mathsfbi{S}_m \boldsymbol{f}(t_0), 
\end{equation}
where
\begin{equation}
\label{eq:comm_Katz}
	\mathsfbi{S}_m = 
	\left(\boldsymbol{I} - \alpha\mathsfbi{A}_m\right)^{-1}
	\cdots
	\left(\boldsymbol{I} - \alpha\mathsfbi{A}_{1}\right)^{-1}
	\left(\boldsymbol{I} - \alpha\mathsfbi{A}_{0}\right)^{-1}
\end{equation}
is referred to as the {\it communicability matrix} for the evolving network.  It provides knowledge of the effective dynamical paths in the direction of time advancement, along which information propagates over the network via the ordered the operations of $\mathsfbi{K}(\mathsfbi{A}_k,\alpha)$.

Similar to the concept of the walk-downweighting parameter $\alpha$, \citet{GrindrodHigham:PRSA2013} considered the use of age-downweighting in the construction of the communicability matrix to further account for the decay of information intensity due to its aging in time.  Their formulation generalizes the product form in (\ref{eq:comm_Katz}) to compute the communicability matrix as
\begin{equation}
\label{eq:comm_gamma}
	\mathsfbi{S}_{k+1} = (\boldsymbol{I} - \alpha\mathsfbi{A}_{k+1})^{- \Delta t_{k+1}} \left[\boldsymbol{I} + e^{-\gamma \Delta t_k} (\mathsfbi{S}_k - \boldsymbol{I})\right],
\end{equation}
with $\mathsfbi{S}_0  = (\boldsymbol{I} - \alpha\mathsfbi{A}_0)^{- \Delta t_0}$ and $\gamma$ being the age-downweighting parameter.  By taking $\Delta t_k = 1$ for all $k$, the communicability matrix $\mathsfbi{S}_m$ computed using equation (\ref{eq:comm_gamma}) recovers the product form in (\ref{eq:comm_Katz}) with $\gamma = 0$ (i.e. infinitely-long memory) and degrades to the instantaneous Katz function $\mathsfbi{K}_m = \left(\boldsymbol{I} - \alpha\mathsfbi{A}_m \right)^{-1}$ with $\gamma \rightarrow \infty$ (i.e. zero memory).  We note that the concept of a tunable memory factor is similar to that adopted in the online dynamic mode decomposition \citep{Zhang:SIAMADS2019}.  Also, when the adjacency matrices $\mathsfbi{A}_k$ are accessed at a constant time interval, we can rescale the time such that $\Delta t_k = 1$ to save the computational cost of $\mathsfbi{S}_m$.

In this study, we use (\ref{eq:comm_gamma}) to compute the communicability matrix $\mathsfbi{S}_m$ that accounts for the network evolution over $t \in [t_0, t_m]$ through the recurrence of $k = 0,1,2,\dots,m$.  Since the operation of the communicability matrix in equation (\ref{eq:io_comm}) resembles that of a state transition operator \citep{SchmidHenningson:2001}, we also consider the SVD of  
\begin{equation}
\label{eq:BMode_SVD}
	\mathsfbi{S}_m \equiv \tilde{\mathsfbi{R}} \tilde{\boldsymbol{\Sigma}} \tilde{\mathsfbi{B}}
\end{equation}
to find the broadcast mode $\tilde{\boldsymbol{b}}$ and receiving mode $\tilde{\boldsymbol{r}}$ by extracting the leading right-singular vector in $\tilde{\mathsfbi{B}}$ and the leading left-singular vector in $\tilde{\mathsfbi{R}}$, respectively.  With $\mathsfbi{S}_m$ tracking all weighted dynamic paths, the mode $\tilde{\boldsymbol{b}}$ gives insights into the gateways of those dynamic paths along which the seeded information can be effectively spread over time.  In what follows, we apply this broadcast mode analysis to the time-evolving network of 2D isotropic turbulence.  Leveraging the shapes of the broadcast modes, we add vortical perturbations to the turbulent flow to explore how these broadcast modes can be used to target the sensitive regions in the turbulence for flow modification.


\section{Application to 2D decaying isotropic turbulence}
\label{sec:setups}

\subsection{Model problem setup}

\begin{figure}
\vspace{+0.02in}
\centering
\begin{tikzpicture}
	\node[anchor=south west,inner sep=0,outer sep=0,fill opacity=0.99] 
		{\includegraphics[width=1.0\textwidth]{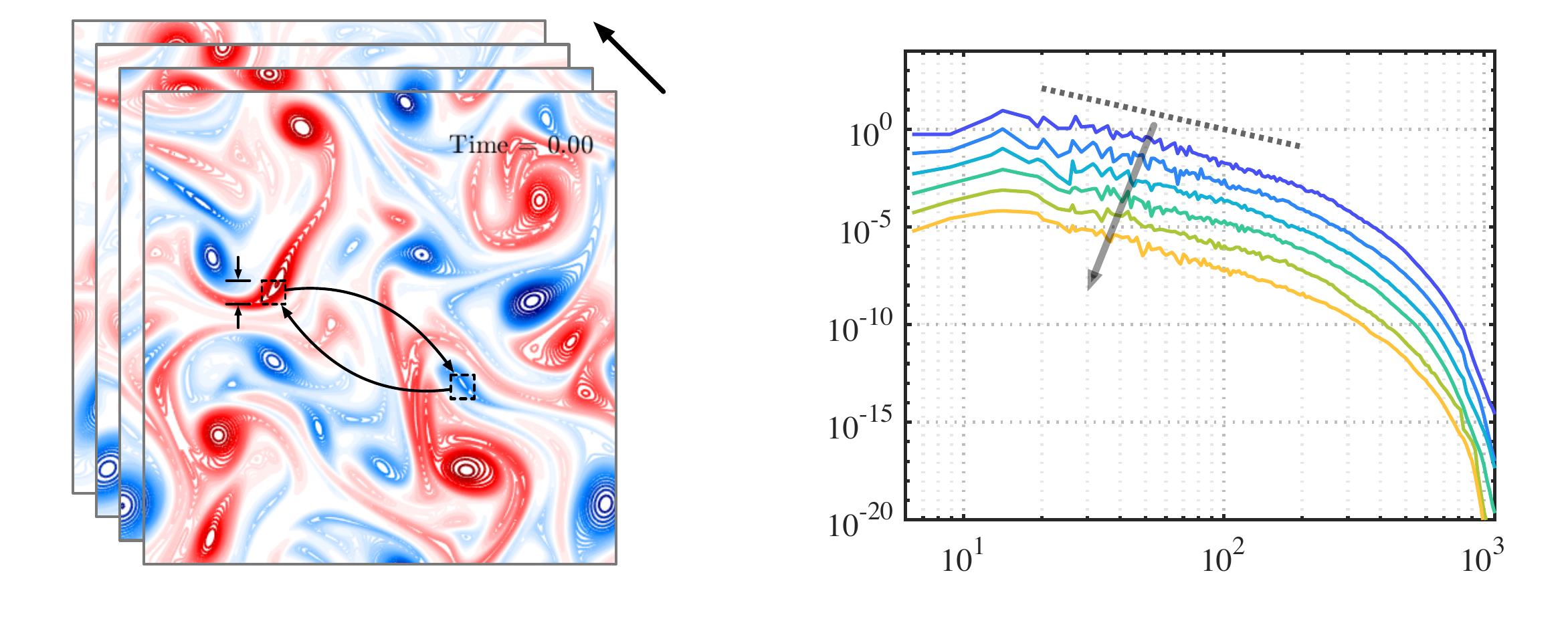}}; 

	\newcommand{\annt}[3]
	{	\FPeval{\xloc}{round(#1*0.01*1.0, 5)}
		\FPeval{\yloc}{round(#2*0.01*1.0, 5)}
		\node[	preaction={fill=white, fill opacity=0.85},
				anchor=south west,
				rounded corners=0.3ex,
				inner sep=0.1ex] 
				at (\xloc\textwidth,\yloc\textwidth) () {#3} ;}
	
	\annt{00}{36}{\indexsize (a)}
	\annt{14}{14}{\rotatebox{90}{\indexsize $\Delta x_n$}}
	\annt{16}{22.4}{\indexsize $\omega(\boldsymbol{x}_i, t)$}
	\annt{28.4}{11.6}{\indexsize $\omega(\boldsymbol{x}_j, t)$}
	\annt{24.8}{20.2}{\indexsize $\mathsfi{A}_{ji}$}	
	\annt{20}{13.2}{\indexsize $\mathsfi{A}_{ij}$}	

	\annt{40.4}{31.5}{\indexsize Time}
	
	\annt{51}{36}{\indexsize (b)}
	\annt{50.0}{19.0}{\rotatebox{90}{\indexsize $E(k_l)$}}
	\annt{68.8}{0}{\indexsize $k_l$ (wavenumber)}
	
	\annt{67}{19}{\scriptsize Time}
	\annt{83.5}{29.3}{\scriptsize $k_l^{-3}$}
	\annt{64.0}{10.0}{\includegraphics[width=0.1in]{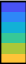}}
	\annt{66.5}{13.2}{\scriptsize$t/t^* = 0$}
	\annt{66.5}{09.4}{\scriptsize$t/t^* = 10.0$}
		
\end{tikzpicture}
\caption{\label{fig:Init_Spec} The 2D isotropic turbulence studied in the present study.  (a) Vorticity field of the initial condition $\omega_0$.  A time-evolving network model of vortical interactions between Eulerian grid points is constructed along the time-trajectory of the turbulent vorticity field. (b) The energy spectra of the time-varying isotropic turbulence with $E(k_l) \sim k_l^{-3}$ power law distribution.  Spectra at different times are vertically shifted by a decade for clarity.}

\end{figure}

We apply the broadcast mode analysis to a 2D decaying isotropic turbulence, which is a complex vortical flow with high levels of unsteadiness.  The time-evolving turbulent flow field analyzed in this study is obtained from the direct numerical simulation (DNS) on a square bi-periodic domain.  The 2D turbulence is simulated by numerically solving the 2D vorticity transport equation
\begin{equation}
\label{eq:vort_eq}
	\partial_t \omega = - \left( \boldsymbol{u} \cdot \boldsymbol{\nabla} \right)\omega + \nu \nabla^2 \omega
\end{equation}
using the Fourier spectral method de-aliased by the $3/2$ rule and the fourth-order Runge--Kutta scheme for time integration \citep{Taira:JFM2016}.  Here, $\omega$ and $\boldsymbol{u}$ denote the vorticity and velocity, respectively, and $\nu$ is the kinematic viscosity.  The simulation is initialized with a vorticity field comprised of a large number of Taylor vortices of random strengths, core sizes, and locations.  We evolve the random vorticity field until the turbulent energy spectrum exhibits the typical power-low profile for 2D isotropic turbulence \citep{Brachet:JFM1988, Benzi:PRA1990, Bracco:PoF2000} and treat this isotropic turbulent flow as the initial condition.  This initial vorticity field and its energy spectra are shown in figure \ref{fig:Init_Spec} (a) and (b), respectively.  Following \citet{Taira:JFM2016} and \citet{Jimenez:JFM2018}, we use this initial flow field to define the characteristic length scale $\lambda \equiv u^* / \omega^*$ and the eddy turnover time $t^* \equiv 1/\omega^*$, where $\omega^* \equiv \langle \omega_0^2 \rangle^{1/2}$ and $u^* \equiv \langle |\boldsymbol{u}_0|^2 \rangle^{1/2}$ are respectively the spatial root-mean-square values for the vorticity and velocity at the initial condition.  Based on this length scale $\lambda$, the Reynolds number chosen in the present study is $Re \equiv u^* \lambda / \nu = 184$, and the size of the square domain is $L = 20 \lambda $, uniformly discretized with $\Delta x = 0.079\lambda$.  The time integration is performed with $\Delta t = 0.0145 t^*$. 

For the construction of the vortical network, we collect vorticity snapshots at a constant time interval $\Delta t_k = 10\Delta t$ over a downsampled mesh with the cell size $\Delta x_n = 2\Delta x$.  We also note that no significant difference was observed in the shapes of the broadcast modes, compared to the downsampling of $\Delta x_n = 4\Delta x$.  The time window for collecting vorticity snapshots is chosen such that the turbulent energy spectra continues to exhibit the power-law profile, as shown in figure \ref{fig:Init_Spec} (b).  In the present study, we consider the Eulerian-based construction of the vortical interaction network \citep{Taira:JFM2016, Murayama:PRE2018, MGM:PRE2018}.  We treat each Cartesian element as a node in the network and the vortical interactions between the elements as the edges, as shown in figure \ref{fig:Init_Spec} (a).   The time-varying nature of the decaying turbulence is characterized by the varying vortical interactions, resulting in a time-evolving network.  By quantifying these interactions, a series of adjacency matrices $\mathsfbi{A}_k$ for vorticity snapshots $\omega(\boldsymbol{x}, t_k)$ can be established, as discussed below.

\subsection{Edge weights: Biot--Savart network and Navier--Stokes network}
We consider two definitions for $\mathsfi{A}_{ij}$ that quantify the interaction between the vortical elements at $\boldsymbol{x}_i$ and $\boldsymbol{x}_j$.  The first one defines the edge weight according to the induced velocity magnitude at $\boldsymbol{x}_i$ imposed by the vortical element at $\boldsymbol{x}_j$ as
\begin{equation}
\label{eq:BS_Net}
	\mathsfi{A}_{ij} = \left| \omega(\boldsymbol{x}_j, t) \right| \Delta x_n^2 \big{/} \left( 2\pi \left|\boldsymbol{x}_i - \boldsymbol{x}_j\right|\right),
\end{equation}
with $\mathsfi{A}_{ii} = 0$ since there is no induced velocity by an vortical element on itself. This Biot--Savart edge weight has been used for network-theoretic models of vortical flows \citep{Nair:JFM2015, Taira:JFM2016, Murayama:PRE2018, MGM:JFM2020}.

The second edge-weight definition considers the change of vorticity evolution at $\boldsymbol{x}_i$ due to a spatial pulse perturbation introduced at $\boldsymbol{x}_j$.  This is computed using
\begin{equation}
\label{eq:NS_Net}
	\mathsfi{A}_{ij} = \left|\left[\mathcal{N}\left(\omega + \epsilon\delta(\boldsymbol{x}_j)\right) - \mathcal{N}\left(\omega\right) \right]_{\boldsymbol{x}_i}\right| \big{/}\epsilon,
\end{equation}
where $\mathcal{N}(\omega) \equiv - \left( \boldsymbol{u} \cdot \boldsymbol{\nabla} \right)\omega + \nu \nabla^2 \omega $ is the right-hand-side of the vorticity transport equation, and $\epsilon\delta(\boldsymbol{x}_j)$ is a vorticity pulse at $\boldsymbol{x}_j$ in the shape of a Taylor vortex
\begin{equation}
	\epsilon\delta(\boldsymbol{x}_j) = \epsilon \left(2/r_\delta - | \boldsymbol{x} - \boldsymbol{x}_j |^2 / r_\delta^3 \right) \exp \left[-| \boldsymbol{x} - \boldsymbol{x}_j |^2 / (2r_\delta^2) \right],
\label{eq:Pulse}
\end{equation} 
with amplitude $\epsilon / u^* = 0.001$ and radius $r_\delta = 1.5\Delta x$.  We note that while a Gaussian-shaped vorticity pulse (i.e. a Lamb--Oseen vortex) can also be considered in equation (\ref{eq:NS_Net}), shaping the pulse as a Taylor vortex ensures the satisfaction of the zero-circulation constraint in the present bi-periodic setting.  Using equation (\ref{eq:NS_Net}), we compute each $\mathsfbi{A}_{ij}$ by evaluating the difference between the two right-hand-side operations at $\boldsymbol{x}_i$, which accounts for the perturbation received at $\boldsymbol{x}_i$ due to the added pulse at $\boldsymbol{x}_j$.  In what follows, we refer to the network defined by the first edge weight as the Biot--Savart network $\mathsfbi{A}^{\text{BS}}$ and the second as the Navier--Stokes network $\mathsfbi{A}^{\text{NS}}$.

\subsection{Downweighting parameters}

With the adjacency matrices constructed, we can determine the Katz function $\mathsfbi{K}$ and the communicability matrix $\mathsfbi{S}_m$ with appropriate choices of the walk- and age-downweighting parameters.  Comparing $\mathsfbi{K} \equiv \left(\boldsymbol{I} - \alpha\mathsfbi{A} \right)^{-1}$ to the resolvent operator generally used in fluid mechanics, the walk-downweighting parameter $\alpha$ can be related to a time scale, or equivalently the inverse of frequency appearing in a resolvent operator \citep{Trefethen:Science1993}.  This is particularly the case for the NS  network, recognizing that the construction of $\mathsfbi{A}^{\text{NS}}$ is closely related to extracting the Jacobian matrix of the vorticity-transport operator about $\omega$.  Moreover, we note that both $\mathsfbi{K}$ and $\mathsfbi{S}_m$ describe the linear spreading of perturbation over a network, as suggested by equations (\ref{eq:Katz_resolvent}) and (\ref{eq:io_comm}).  Based on these observations, we seek the insights from the rapid distortion theory to choose the values of these downweighting parameters.

The rapid distortion theory was developed based on the linearized vorticity transport equation, with the attempt to describe the evolution of turbulence \citep{Batchelor:QJMAM1954}.  Through scaling arguments, \citet{Hunt:JFM1990} have shown that rapid distortion theory is generally valid within a time-horizon shorter than a unit eddy turnover time.  Therefore, we utilize the eddy turnover time $t^*$ to choose the age-downweighting parameter $\gamma = 1/t^*$ and set the value of the walk-downweighting parameter $\alpha = e^{-\Delta t_k/t^*}$ such that the information intensity downweighted by a walk within the time slice $t_k$ is equal to that due to the march to the next time slice $t_{k+1}$, as in (\ref{eq:comm_gamma}).  We also note that the concept of choosing a finite time-horizon is adopted in a similar fashion in the discounted resolvent analysis \citep{Jovanovic:Thesis2004, Yeh:JFM2019, Yeh:PRF2020}.  This choice of $\alpha$ also satisfies $\alpha < 1/\rho(\mathsfbi{A}_k)$ for all $k$, allowing for the Katz function to reduce to the resolvent form.  With the broadcast mode extracted through the SVD in (\ref{eq:BMode_SVD}), we investigate its use for modifying the evolution of the turbulent flow.

\subsection{Broadcast-mode-based perturbation}
Since the broadcast mode $b(\boldsymbol{x})$ identifies effective nodes to initialize perturbations, we use it to perturb the initial condition of the turbulent vorticity field as
\begin{equation}
\label{eq:PertInit}
	\omega_p(\boldsymbol{x}, t_0) = \omega_0(\boldsymbol{x}) + a \left[ b(\boldsymbol{x}) - \langle b(\boldsymbol{x}) \rangle \right].
\end{equation} 
Here, the removal of the spatial mean $\langle b(\boldsymbol{x}) \rangle$ ensures that no net circulation is introduced to the bi-periodic domain due to the added perturbation \citep{Jimenez:JFM2018}.  The perturbation amplitude is chosen such that $a||b - \langle b \rangle ||_2/||\omega_0||_2 = 0.001$.  We evolve the turbulent flow from this perturbed initial condition $\omega_p(\boldsymbol{x}, t_0)$ and track the flow modification \citep{Jimenez:JFM2018,Jimenez:JT2020} with
\begin{equation}
\label{eq:DeltaQ}
	\Delta \omega(\boldsymbol{x}, t) = \omega_p(\boldsymbol{x}, t) - \omega(\boldsymbol{x}, t).
\end{equation} 
This modification is assessed with respect to the broadcast modes extracted from the NS and BS networks and to different time horizons $t_m$ over which the broadcast modes are extracted.

\section{Results}
\label{sec:results}

\subsection{Biot--Savart vs. Navier--Stokes broadcast modes}

\begin{figure}
\vspace{0.20in}
\centering
\begin{overpic}[width=1.0\textwidth]{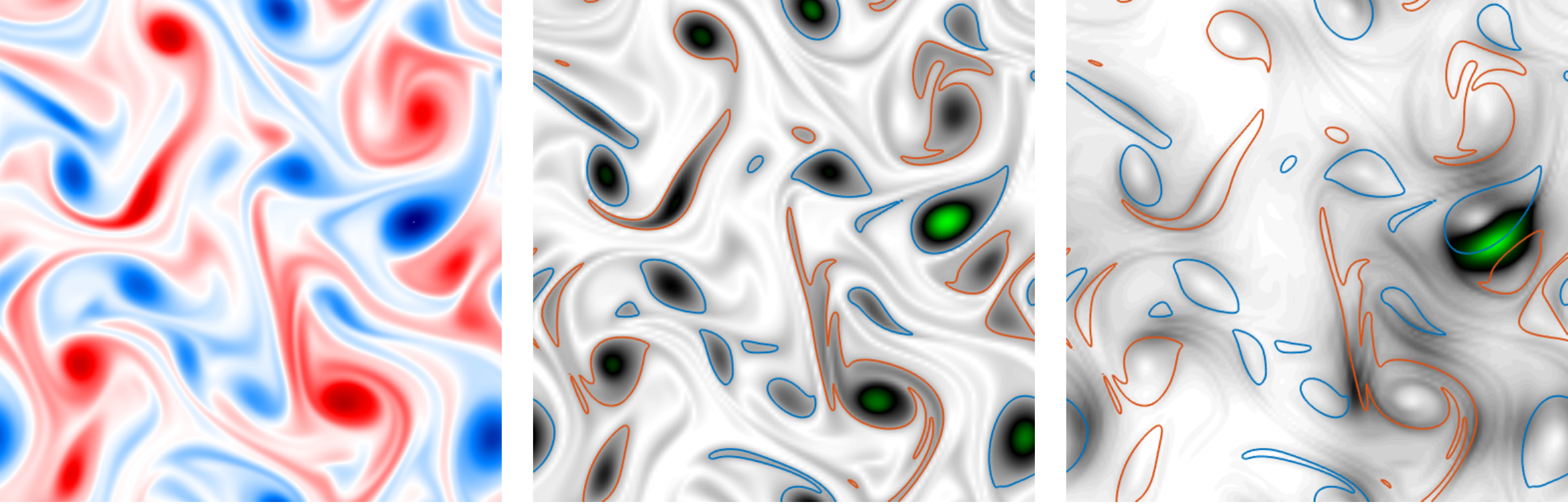}
	\put(04.0, 33.5){\indexsize (a) Vorticity field $\omega_0/\omega^*$}
	\put(36.0, 33.5){\indexsize (b) BS broadcast mode ($\boldsymbol{b}^{\text{BS}}$)}
	\put(70.3, 33.5){\indexsize (c) NS broadcast mode ($\boldsymbol{b}^{\text{NS}}$)}
	\put(05.0, 26.0){\includegraphics[width=0.1in]{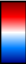}}
	\put(02.1, 26.0){\scriptsize$-5$}
	\put(03.8, 29.5){\scriptsize$5$}
	
	\put(39.0, 26.0){\includegraphics[width=0.1in]{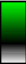}}
	\put(37.7, 26.0){\scriptsize$0$}
	\put(34.9, 29.5){\scriptsize$0.03$}
	
	\put(73.0, 26.0){\includegraphics[width=0.1in]{CBar_BW}}
	\put(71.7, 26.0){\scriptsize$0$}
	\put(68.9, 29.5){\scriptsize$0.04$}
\end{overpic}
\caption{\label{fig:NS_BS_BModes}
The broadcast modes of BS and NS networks. (a) The vorticity field about which the broadcast modes are computed; (b) the BS broadcast mode ($\boldsymbol{b}^{\text{BS}}$); and (c) the NS broadcast mode ($\boldsymbol{b}^{\text{NS}}$). Vorticity contours for $\omega/\omega^* = \pm 1$ are superposed on the modes in (b-c) for comparison.}
\end{figure}

For the initial vorticity field shown in figure \ref{fig:NS_BS_BModes}(a), we extract the broadcast mode from $\mathsfbi{S}_0  = (\boldsymbol{I} - \alpha\mathsfbi{A}_0)^{- \Delta t_0}$ using equation (\ref{eq:BMode_SVD}).  The broadcast modes for the BS and NS networks are presented in figure \ref{fig:NS_BS_BModes}(b) and (c), respectively as  $\boldsymbol{b}^{\text{BS}}$ and  $\boldsymbol{b}^{\text{NS}}$.   The BS broadcast mode suggests that the vortex cores, featured with high levels of vorticity, are the regions of high broadcast strength.  The study of \citet{Taira:JFM2016} identified these regions as the network `hubs', which are characterized by high levels of vortical interaction occurring in turbulent flow.  In a similar problem setting, \citet{Jimenez:JFM2018} also labelled the vortices as the influential structures that dominate the evolution of the 2D turbulence. 

The NS broadcast mode paints a different picture. We find that the volumes of high broadcast strength occupy the regions between opposite-sign vortex pairs, as highlighted in the magenta boxes in figure \ref{fig:NS_BS_BModes}(c).  This observation agrees with \citet{Jimenez:JT2020}, where these vortex `dipoles' are identified as the influential structures in the 2D turbulence through a Monte--Carlo based search over all subvolumes in the flow.  These vortex dipoles, acting as local `jets,' locally build up shear layers that are sensitive to perturbations.  Moreover, contrary to the BS broadcast mode, the NS broadcast mode reveals that the large vortex cores are the regions of the lowest broadcast strength.  In 2D unforced turbulence, the only mechanism for vortex deformation is through a strain field of comparable strength to the vortex \citep{PullinSaffman:ARFM1998}.  For large isolated vortices in 2D isotropic turbulence, such deformations are rare unless they form dipoles or merge with one another \citep{McWilliams:JFM1990}.  This is reflected in the NS edge weight (\ref{eq:NS_Net}), which can be expanded as $\mathsfi{A}_{ij} \propto \boldsymbol{u} \cdot \bnabla \delta\omega + \delta \boldsymbol{u} \cdot \bnabla \omega$ with $\delta \omega$ and $\delta \boldsymbol{u}$ being the introduced vorticity and velocity by the pulse perturbation, respectively.  Here, we do not include the viscous term $\nu \nabla^2 \delta \omega $ since it is equal for all nodes and contributes no significant difference to the centrality measure.  When the small vorticity pulse is introduced at the core of large vortices, the induced $\bnabla \delta \omega$ and $\delta \boldsymbol{u}$ are respectively perpendicular to the local base flow $\boldsymbol{u}$ and $\bnabla \omega$, resulting in weak perturbations to all vortical elements in the 2D turbulence and hence resulting in the low broadcast strength.  

\subsection{Flow modification with initial perturbation and the receive modes}

\begin{figure}
\label{fig:DeltaQ_Time}
\centering
\begin{overpic}[scale=0.6]{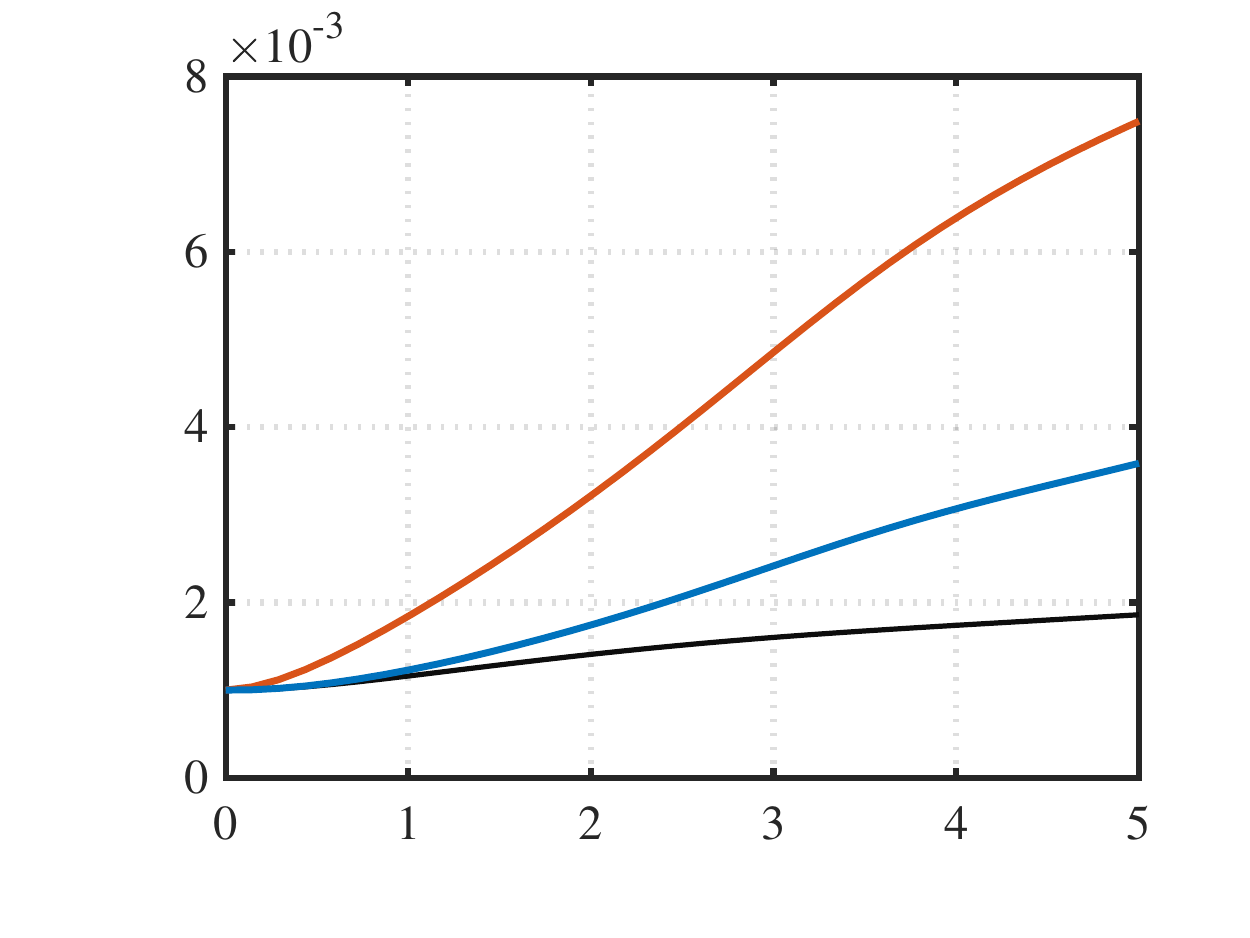}
	\put(07.0, 28.0){\rotatebox{90}{\indexsize $||\Delta\omega||_2/||\omega_0||_2$}}
	\put(53.0, 03.0){\indexsize $t/t^*$}
	\put(53.0, 19.7){\indexsize random perturbation}
	\put(62.0, 32.5){\indexsize\color{blue}$\boldsymbol{b}^{\text{\tiny BS}}$}
	\put(58.0, 49.5){\indexsize\color{red}$\boldsymbol{b}^{\text{\tiny NS}}$}
\end{overpic}
\caption{\label{fig:DelQ_Init} Time evolution of flow modification $\Delta\omega$ achieved by NS broadcast-mode perturbation ({\color{red}$\boldsymbol{b}^{\text{\tiny NS}}$}), BS broadcast-mode perturbation ({\color{blue}$\boldsymbol{b}^{\text{\tiny BS}}$}), and Taylor vortices of random locations, sizes and strengths.}
\end{figure}
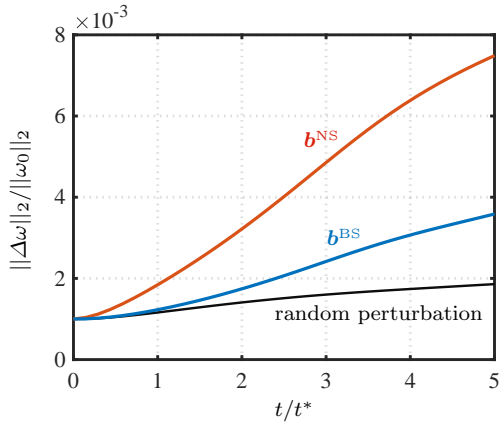

\begin{figure}
\label{fig:DeltaQ_BSNS}
\vspace{0.2in}
\begin{overpic}[width=0.99\textwidth]{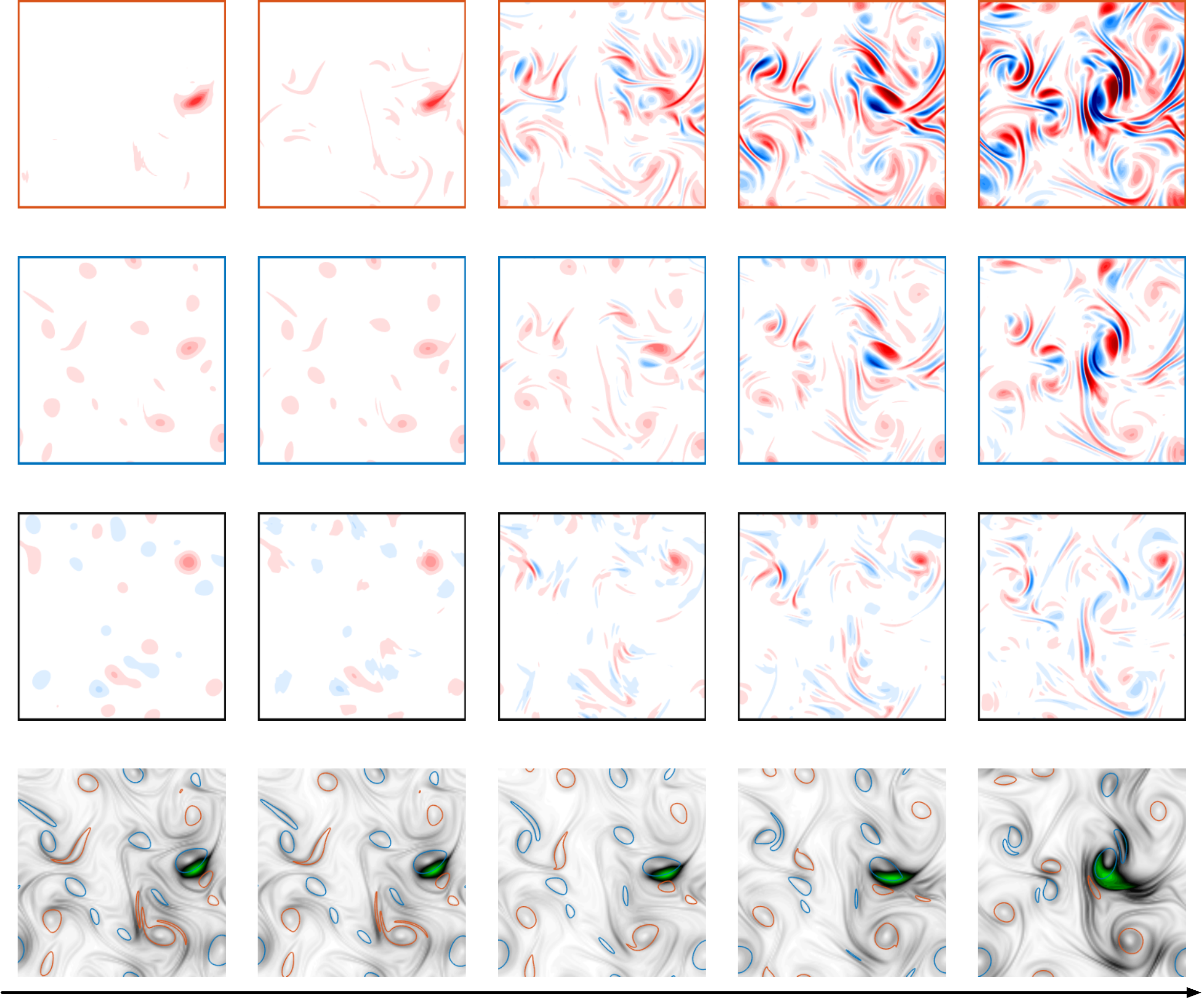}
	\put( 1.0, 84.0 ){\indexsize (a) $\Delta\omega/\omega^*$ for {\color{red}$\boldsymbol{b}^{\text{\tiny NS}}$ perturbation}}
	\put( 1.0, 62.75){\indexsize (b) $\Delta\omega/\omega^*$ for {\color{blue}$\boldsymbol{b}^{\text{\tiny BS}}$ perturbation}}
	\put( 1.0, 41.5 ){\indexsize (c) $\Delta\omega/\omega^*$ for random perturbation}
	\put( 1.0, 20.25){\indexsize (d) NS receiving modes $\boldsymbol{r}^{\text{\tiny NS}}(t)$}
	
	\put(02.5, 67.5){\includegraphics[width=0.07in]{CBar_BR}}
	\put(02.5, 72.0){\tiny$\Delta\omega/\omega^*$}
	\put(04.0, 70.3){\tiny$+0.03$}
	\put(03.9, 67.4){\tiny$-0.03$}
	
	\put( 6.0, -2.0){\indexsize $t/t^* = 0$}
	\put(29.0, -2.0){\indexsize $0.29$}
	\put(49.0, -2.0){\indexsize $1.45$}
	\put(69.0, -2.0){\indexsize $2.90$}
	\put(88.5, -2.0){\indexsize $4.93$}
	
	\put(02.5, 03.5){\includegraphics[width=0.07in]{CBar_BW}}
	\put(02.5, 08.0){\tiny$\boldsymbol{r}^{\text{\tiny NS}}$}
	\put(04.2, 03.4){\tiny$0$}
	\put(04.2, 06.3){\tiny$0.05$}
\end{overpic}
\vspace{0.1in}
\caption{\label{fig:DelQField_Init} Visualization of $\Delta\omega(\boldsymbol{x})$ achieved by the initial perturbations in the shape of (a) NS broadcast mode {\color{red}$\boldsymbol{b}^{\text{\tiny NS}}$}, (b) BS broadcast mode {\color{blue}$\boldsymbol{b}^{\text{\tiny BS}}$}, and (c) random Taylor vortices. (d) The NS receiving modes extracted from the communicability matrix $\mathsfbi{S}_m$ constructed over the time horizon $[0, t]$, with vorticity contours for $\omega/\omega^* = \pm 1.5$ superposed on the receiving modes for comparison.}
\end{figure}

Here, we perturb the initial vorticity field using $\boldsymbol{b}^{\text{BS}}$ and $\boldsymbol{b}^{\text{NS}}$ according to equation (\ref{eq:PertInit}) and track in time the modifications of the turbulent flow with $\Delta\omega$.  The modifications achieved by the broadcast modes are shown in figure \ref{fig:DelQ_Init} and compared to that by a perturbation of randomly superposed Taylor vortices, which results in low levels of $\Delta\omega$ over time.  While the perturbations of both broadcast modes show much better capability for flow modification than random perturbation, the NS broadcast mode produces even higher $\Delta \omega$ than that made by the BS broadcast mode.  The flow modifications $\Delta\omega(\boldsymbol{x}, t)$ are visualized in figure \ref{fig:DelQField_Init} (a--c).  We observe that the modification achieved by the $\boldsymbol{b}^{\text{BS}}$ perturbation remains in the vortex cores shortly after the initial condition at $t/t^*=0.29$.  At the same time, the perturbation based on $\boldsymbol{b}^{\text{NS}}$ has lost its initial shape due to the interactions with turbulence.  Evolving in time, the modification achieved by $\boldsymbol{b}^{\text{NS}}$ perturbation spreads over space with amplifying magnitude, while in the $\boldsymbol{b}^{\text{BS}}$ case the regions of high $\Delta\omega$ remain in the center of the domain at even $t/t^* = 4.93$.  

We also note that even though the shapes of the initial perturbations and the levels of $||\Delta \omega||_2$ are different for all the three cases, the structures of $\Delta \omega(\boldsymbol{x})$ share similar signatures, particularly at the later times of $t/t^* = 2.90$ and $4.93$ and around the main vortex dipole which gradually moves from the right to the center of the domain.  This observation motivates us to further examine the receiving modes, which we show in figure \ref{fig:DelQField_Init} (d).  These NS receiving modes $\boldsymbol{r}^{\text{NS}}$ are extracted from $\mathsfbi{S}_m$ using equation (\ref{eq:BMode_SVD}), with $t_m$ being the same instant at which $\Delta\omega(\boldsymbol{x})$ are visualized in figure \ref{fig:DelQField_Init} (a--c).  We find agreements between the structures of $\Delta \omega$ and $\boldsymbol{r}^{\text{NS}}$, particularly in the region of the main vortex dipole and the pattern of the streaks.  According to $\boldsymbol{r}^{\text{NS}}$, the main vortex dipole is also the most receptive structure to perturbations, in addition to its high broadcasting capability.  This shows that the vortex dipole is not only the main driver for spreading and amplifying perturbation, but also a highly responsive structure to perturbations, forming an internal feedback loop to continuously amplify existing perturbations.  

\subsection{Time-evolving broadcast mode: effects of time horizon}

\begin{figure}
\label{fig:DeltaQ_BSNS}
\vspace{0.25in}
\centering
\begin{overpic}[width=0.99\textwidth]{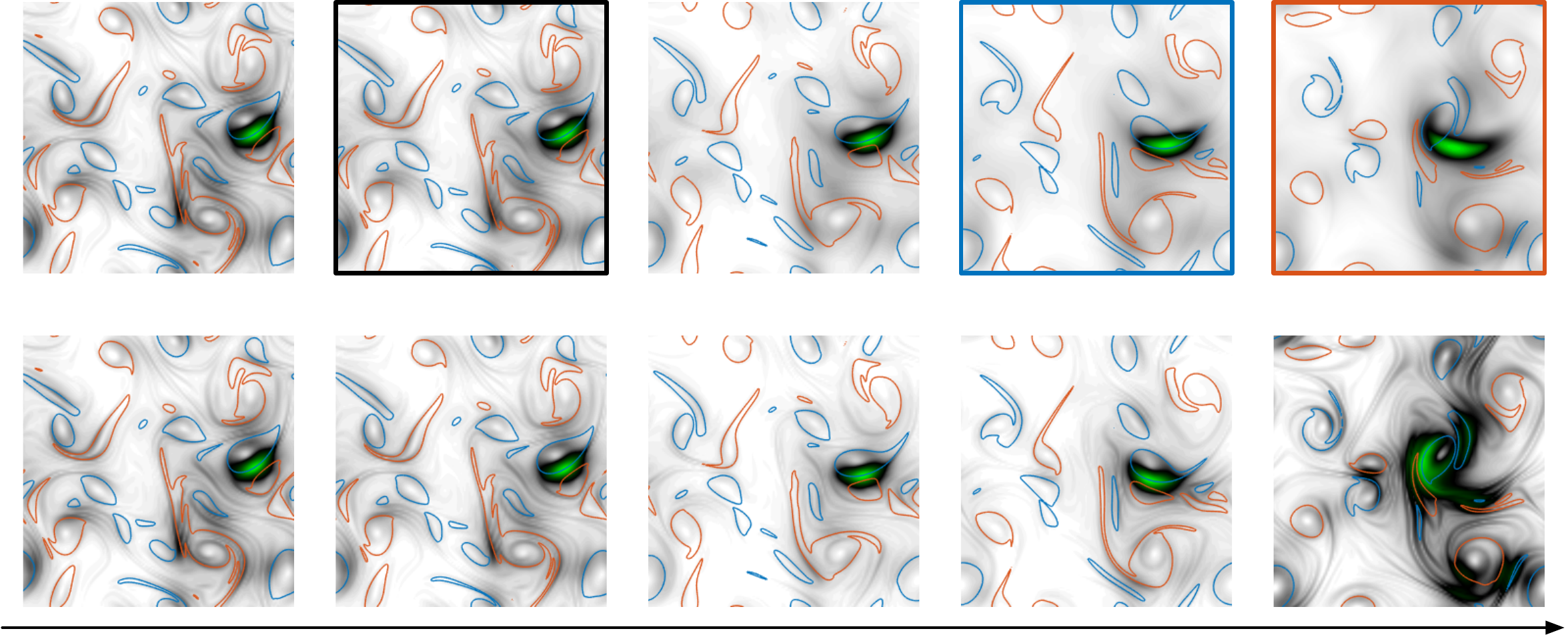}
	\put( 1.0, 42.0 ){\indexsize (a) $\boldsymbol{b}^{\text{\tiny NS}}(t_m)$ with finite-time memory ($\gamma = 1/t^*$)}
	\put( 1.0, 20.5 ){\indexsize (b) $\boldsymbol{b}^{\text{\tiny NS}}(t_m)$ with zero memory ($\gamma \rightarrow \infty$)}

	\put(02.5, 24.5){\includegraphics[width=0.07in]{CBar_BW}}
	\put(04.2, 24.4){\tiny$0$}
	\put(04.2, 27.3){\tiny$0.05$}

	\put( 6.0, -2.0){\indexsize $t_m/t^* = 0$}
	\put(29.0, -2.0){\indexsize $0.29$}
	\put(49.0, -2.0){\indexsize $1.45$}
	\put(69.0, -2.0){\indexsize $2.90$}
	\put(88.5, -2.0){\indexsize $4.93$}
\end{overpic}
\vspace{0.1in}
\caption{\label{fig:BMode_vs_Te} 
Broadcast modes $\boldsymbol{b}(t_m)$ for the evolving networks over different time horizons $t \in [0, t_m]$ using (a) finite-time memory with $\gamma = 1/t^*$ and  (b) zero memory with $\gamma \rightarrow \infty$.  Note that in the latter case the evolving network degrades to a time-frozen one at the same instant of $t_m$, as the communicability matrix becomes the instantaneous Katz function.  The broadcast modes used here are for $\mathsfbi{A}^{\text{NS}}$. Instantaneous vorticity fields are superposed on broadcast modes with representative contour lines.  The flow modifications achieved by the initial perturbations shaped in finite-memory broadcast modes of $t_m/t^* = 0.29$, $2.90$ and $4.93$ are shown in figure \ref{fig:DelQ_TM} correspondingly.}
\end{figure}

Next, we examine the effect of the time horizon on the structure of the broadcast modes and their capability of modifying the turbulent flow.  Here, we focus on the NS broadcast modes, since they are more effective in modifying the turbulent flow compared to the BS broadcast mode in the present setting.  In figure \ref{fig:BMode_vs_Te}, we show the broadcast modes extracted from the communicability matrices of finite-time memory $\gamma = 1/t^*$ and zero memory $\gamma \rightarrow \infty$.  Note that in the latter case, the communicability matrix $\mathsfbi{S}_m$ reduces to the instantaneous Katz function, $\mathsfbi{K}_m = \left(\boldsymbol{I} - \alpha\mathsfbi{A}_m \right)^{-1}$. 

\begin{figure}
\centering
\begin{overpic}[scale=0.6]{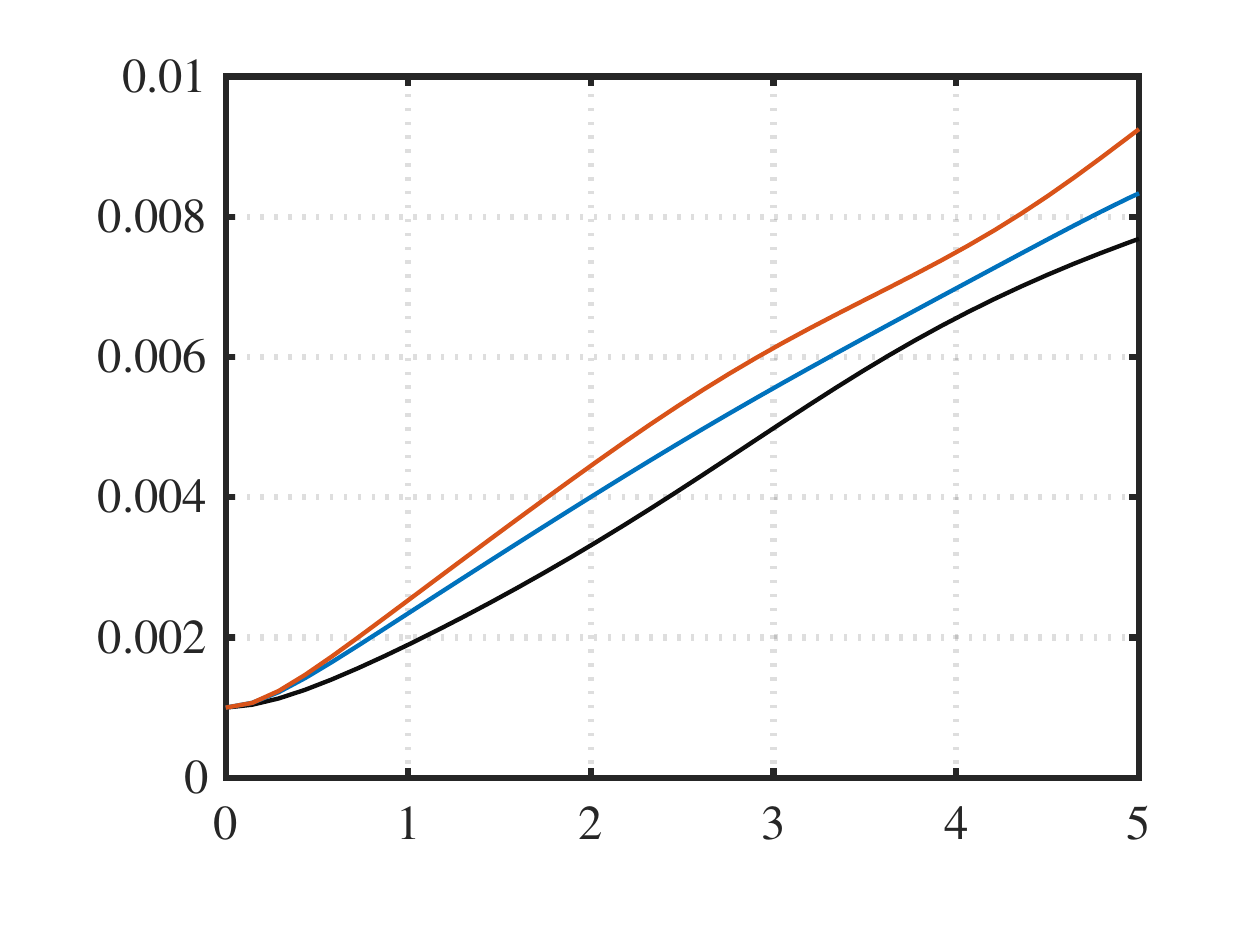}
	\put(02.0, 28.0){\rotatebox{90}{\indexsize $||\Delta\omega||_2/||\omega_0||_2$}}
	\put(53.0, 03.0){\indexsize $t/t^*$}
	\put(55.0, 30.0){\scriptsize\color{red}$t_m/t^* = 4.93$}
	\put(55.0, 25.0){\scriptsize\color{blue}$t_m/t^* = 2.90$}
	\put(55.0, 20.0){\scriptsize $t_m/t^* = 0.29$}
\end{overpic}
\caption{\label{fig:DelQ_TM} Flow modifications $\Delta\omega$ via initial perturbations shaped by the NS broadcast-mode extracted from $\mathsfbi{S}_m$ over different time horizons.}
\end{figure}
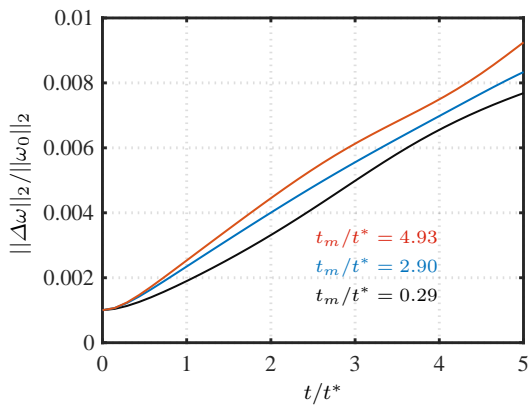

We vary the time horizon $t \in [0, t_m]$ over which the communicability matrices are constructed to examine its effect on the shape of the extracted broadcast modes.  Compared to the instantaneous broadcast modes found from the time-frozen network using (\ref{eq:Katz_SVD}), the broadcast modes for the evolving network model show the reminiscence of the path along which the primary vortex dipole passes, as depicted by the instantaneous vorticity contours overlaid on the broadcast modes.  This is evident for the modes at $t_m/t^* = 2.2$ and $4.5$, and is attributed to the memory effect of the time-evolving model.  In figure \ref{fig:DelQ_TM},  we show the capabilities to modify the turbulent flow using the initial perturbations shaped by these finite time-horizon broadcast modes.  We find that, with longer time horizon, higher level of flow modification is achieved by the corresponding broadcast-mode-based initial perturbation.  The observations we made from the broadcast modes and receiving modes show the potential held by the time-evolving network model for investigating the input--output process for a time-varying base flow.  It gives guidance to the correct gateways to the effective dynamical path along which perturbations can grow and also captures the flow responses to the existing perturbations.

We have examined the capability of broadcast-mode-based perturbation to modify the evolution of turbulent flow.  To this point, the modification is assessed according to the change in the flow field without a target state.  Next, we explore the use of the broadcast mode for modifying the flow along a prescribed direction.

\subsection{Feedforward control for accelerating and decelerating energy dissipation}

In this section, we perform the broadcast mode analysis in real-time with the flow simulation with the goal of modifying the flow by accelerating or decelerating the dissipation of kinetic energy.  To modify the flow in a prescribed direction, we consider the signed NS edge weight,
\begin{equation}
\label{eq:NS_Net_Signed}
		\mathsfi{A}_{ij} = \text{sign}\left(\omega(\boldsymbol{x}_i)\right) \left[~\mathcal{N}(\omega + \epsilon\delta(\boldsymbol{x}_j)) - \mathcal{N}(\omega)~\right]_{\boldsymbol{x}_i} /\epsilon,
\end{equation}
to account for the alignment in the directions of the base flow $\omega(\boldsymbol{x}_i)$ and the perturbation received at $\boldsymbol{x}_i$.  
Hence, a positive-valued $\mathsfi{A}_{ij}$ implies that a favorable perturbation to the base flow $\omega(\boldsymbol{x}_i)$ is received at $\boldsymbol{x}_i$ when introducing a vorticity pulse in the sign of $\epsilon$ at $\boldsymbol{x}_j$.  We can set this perturbation at $\boldsymbol{x}_i$ into an opposing effort by reversing the sign of $\epsilon$ for the vorticity perturbation at $\boldsymbol{x}_j$, since (\ref{eq:NS_Net_Signed}) is linear for sufficiently small $\epsilon$.

\begin{figure}
\vspace{0.1in}
\centering
\begin{tikzpicture}
	\node[anchor=south west,inner sep=0,outer sep=0,fill opacity=1.0] 
		{\includegraphics[width=0.99\textwidth]{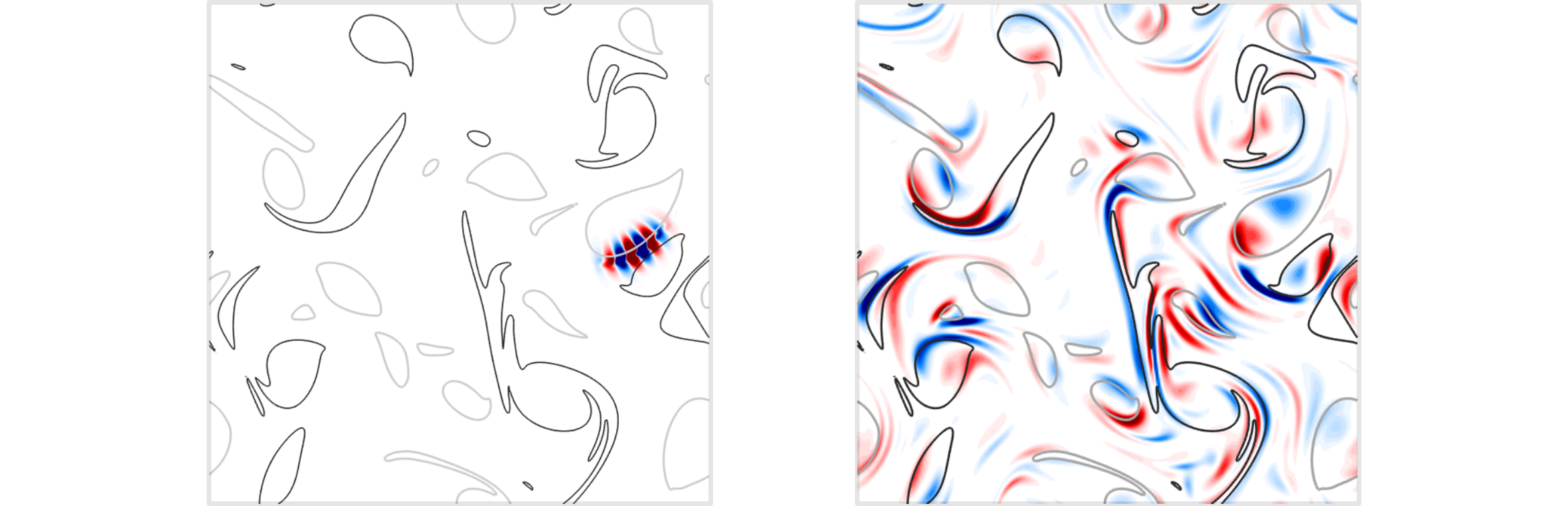}}; 

	\newcommand{\annt}[3]
	{	\FPeval{\xloc}{round(#1*0.01*0.99, 5)}
		\FPeval{\yloc}{round(#2*0.01*0.99, 5)}
		\node[	preaction={fill=white, fill opacity=0.0},
				anchor=south west,
				rounded corners=0.3ex,
				inner sep=0.1ex] 
				at (\xloc\textwidth,\yloc\textwidth) () {#3} ;}
	\newcommand{\anns}[3]
	{	\FPeval{\xloc}{round(#1*0.01*0.99, 5)}
		\FPeval{\yloc}{round(#2*0.01*0.99, 5)}
		\node[	preaction={fill=white, fill opacity=0.9},
				anchor=south west,
				rounded corners=0.3ex,
				inner sep=0.1ex] 
				at (\xloc\textwidth,\yloc\textwidth) () {#3} ;}

	\anns{15}{29}{\indexsize Mode $\boldsymbol{b}_1$}
	\annt{15}{23}{\includegraphics[width=0.1in]{CBar_BR}}
	\annt{17.2}{22.6}{\scriptsize$-0.05$}
	\annt{17.2}{26.5}{\scriptsize$+0.05$}
	
	\anns{56}{29}{\indexsize Mode $\boldsymbol{b}_4$}
	\annt{56}{23}{\includegraphics[width=0.1in]{CBar_BR}}
	\annt{58.2}{22.6}{\scriptsize$-0.02$}
	\annt{58.2}{26.5}{\scriptsize$+0.02$}
\end{tikzpicture}
\caption{\label{fig:BMode_CloseLopp} Broadcast modes extracted from the signed N-S network, overlaid with the instantaneous vorticity field (black/gray). }
\vspace{-0.15in}
\end{figure}

To perform feedforward control, for the instantaneous vorticity field we extract the broadcast modes from the Katz function using (\ref{eq:Katz_SVD}).  Since the goal here is to modify the flow in a prescribed direction rather than simply achieving a high-level modification, we consider the broadcast mode $\boldsymbol{b}$ associated with the highest gain-scaled mean value of the receive mode $\sigma\left|\langle\boldsymbol{r}\rangle\right|$, where $\sigma$ is the corresponding singular value.  With the definition of the signed edge, this $\boldsymbol{b}$ generates the response $\boldsymbol{r}$ of the highest global alignment to the direction of the base flow.  The broadcast modes $\boldsymbol{b}_1$ and $\boldsymbol{b}_4$ for the signed NS edges are shown in figure \ref{fig:BMode_CloseLopp}.  Similar to the observations made for the unsigned edges, the leading broadcast mode $\boldsymbol{b}_1$ reveals the sensitive region in the vortex dipole, but further identifies a wave-like structure with the use of the signed edge.   To modify the flow by acceleration or decelerating the energy dissipation, we adopt the subdominant broadcast mode $\boldsymbol{b}_4$ with the highest $\sigma_4\left|\langle\boldsymbol{r}_4\rangle\right|$, which shows that the streaks occupying the zero-vorticity regions are the sensitive structures.  We also note that, although this mode is extracted via the SVD, it is possible to be casted into an optimization problem that maximizes $\sigma\left|\langle\boldsymbol{r}\rangle\right|$.  

\begin{figure}
\vspace{0.1in}
\centering
\begin{tikzpicture}
	\node[anchor=south west,inner sep=0,outer sep=0,fill opacity=1.0] 
		{\includegraphics[width=0.99\textwidth]{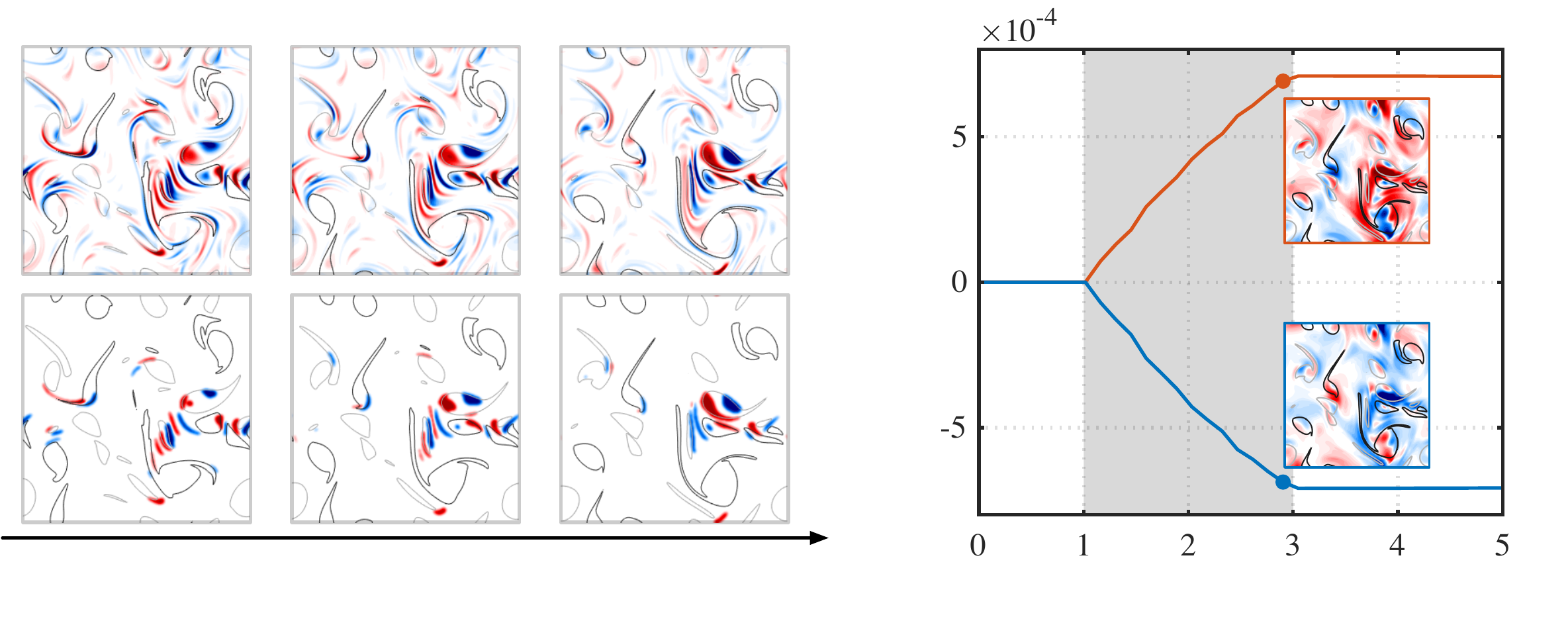}}; 

	\newcommand{\annt}[3]
	{	\FPeval{\xloc}{round(#1*0.01*0.99, 5)}
		\FPeval{\yloc}{round(#2*0.01*0.99, 5)}
		\node[	preaction={fill=white, fill opacity=0.0},
				anchor=south west,
				rounded corners=0.3ex,
				inner sep=0.1ex] 
				at (\xloc\textwidth,\yloc\textwidth) () {#3} ;}
	\newcommand{\anns}[3]
	{	\FPeval{\xloc}{round(#1*0.01*0.99, 5)}
		\FPeval{\yloc}{round(#2*0.01*0.99, 5)}
		\node[	preaction={fill=white, fill opacity=0.9},
				anchor=south west,
				rounded corners=0.3ex,
				inner sep=0.1ex] 
				at (\xloc\textwidth,\yloc\textwidth) () {#3} ;}
	
	\annt{0}{37}{\indexsize (a)}
	\annt{03.5}{01.5}{\indexsize $t/t^* = 1.16$}
	\annt{20.5}{01.5}{\indexsize $t/t^* = 2.03$}
	\annt{37.5}{01.5}{\indexsize $t/t^* = 2.90$}
	\annt{06.5}{32.0}{\includegraphics[width=0.07in]{CBar_BR}}
	\anns{01.5}{31.7}{\tiny$-0.04$}
	\anns{03.0}{34.7}{\tiny$0.04$}

	\annt{06.5}{16.2}{\includegraphics[width=0.07in]{CBar_BR}}
	\anns{01.5}{15.9}{\tiny$-0.05$}
	\anns{03.0}{18.9}{\tiny$0.05$}
		
	\annt{56}{37}{\indexsize (b)}
	\annt{57}{17}{\rotatebox{90}{\indexsize $\Delta E / E_0$}}
	\annt{78}{01}{\indexsize $t/t^*$}
	\annt{71}{21}{\scriptsize control on}
	\anns{83.5}{32.5}{\scriptsize\color{red} Case $+a$}
	\anns{83.5}{09}{\scriptsize\color{blue} Case $-a$}	

	\annt{89.5}{19.6}{\includegraphics[width=0.07in]{CBar_BR}}
	\anns{84.5}{19.3}{\tiny$-0.05$}
	\anns{86.0}{22.3}{\tiny$0.05$}
\end{tikzpicture}
\caption{\label{fig:deltaTKE_FFWCtrl} (a) Broadcast modes extracted from the signed NS network (top) and the designed forcing based on the top $1\%$ of the nodes (bottom), overlaid with the instantaneous vorticity field (black/gray). (b) The change in total kinetic energy over the domain $\Delta E(t) = \int_\mathcal{D} \Delta e(\boldsymbol{x}, t) {\rm d}\boldsymbol{x}$ normalized by the initial kinetic energy $E_0 = \int_\mathcal{D} e_0(\boldsymbol{x}, t_0) {\rm d}\boldsymbol{x}$.  Inserted are the flow modification in terms of turbulent kinetic energy $\Delta e(\boldsymbol{x})/ \langle |\boldsymbol{u}_0|^2 \rangle$ at $t/t^* = 2.90$.   }
\end{figure} 

We design the broadcast-mode-based forcing $\boldsymbol{b}_f$ for feedforward control by adding Gaussian vorticity pulses to the nodes in the top $1\%$ of positive and negative broadcast strength, ensuring zero circulation in the right-hand-side forcing.  The feedforward control is performed by adding a right-hand-side forcing of 
\begin{equation}
	\partial_t\omega = \mathcal{N}(\omega) + a\boldsymbol{b}_f(\omega), 
\end{equation}
with the forcing amplitude $a\,t^* || \boldsymbol{b}_f ||_2 / || \omega_0 ||_2 = 0.001$.  The instantaneous broadcast mode of the highest gain-scaled mean value and the resulting forcing shape are shown in figure \ref{fig:deltaTKE_FFWCtrl} (a).  We turn on the control over $t/t^* \in [1, 3]$ with positive forcing amplitude (case $+a$) in favor to the base flow direction and with negative amplitude (case $-a$) in the opposite way, while tracking the change in kinetic energy $\Delta e(\boldsymbol{x}) = || \boldsymbol{u}_\text{ctrl}(\boldsymbol{x}) ||_2^2 - || \boldsymbol{u}_\text{unctrl}(\boldsymbol{x}) ||_2^2$.  The histories of the changes in the total kinetic energy and the instantaneous $\Delta e(\boldsymbol{x})$ at $t/t^* = 2.90$ are shown in figure \ref{fig:deltaTKE_FFWCtrl} (b).  We find that the broadcast-mode-based control is able to modify the turbulent flow in the prescribed direction for both cases.  The signed  signed broadcast modes can be utilized to pin down the effective flow regions and introduce forcing for either accelerating or decelerating energy dissipation, showing the potential of the present broadcast analysis for control of turbulent flows.

\section{Conclusion}
\label{sec:conclusion}

We introduced the broadcast analysis by blending the resolvent analysis and the network Katz centrality to identify the influential structures for time-varying base flows.  We applied the analysis to 2D decaying isotropic turbulence to reveal the principal structures that can effectively modify turbulent flow. Due to the time-varying nature of 2D turbulence, we model it as an evolving network of vortical elements.  By blending the formulations of Katz centrality and resolvent analysis, the broadcast analysis performs the SVD of the Katz function and the communicability matrix.  The broadcast modes are extracted from the leading right-singular vectors, enabling us to pinpoint the vortical elements that effectively spread perturbations over the evolving turbulent vortical network.  

For the vortical network, we considered two definitions for the edge weights to quantify the strength of the vortical interactions, yielding the BS and NS broadcast modes.  We leveraged the insights given by the broadcast modes to design distributive forcing input for the turbulent flow with the goal of modifying its time evolution.  While both BS and NS broadcast modes are shown to be able to effectively modify the flow evolution, we observe that the N--S broadcasting-mode-based perturbation is more effective in modifying flow evolution by focusing the actuation efforts towards the vortex dipoles.  By constructing the communicability matrix over a longer time horizon to account for the longer history of the turbulent evolution, we show that the extracted broadcast modes are able to achieve higher levels of flow modification.  The broadcast mode can also be extracted from a signed network, guiding flow modification in a desired direction in a feedforward control setup.  The present network-inspired approach serves as a novel tool to analyze time-varying turbulent base flows and provide flow control guidance.

\section*{Acknowledgments}
\indent We gratefully acknowledge the support from the Office of Naval Research (N00014-19-1-2460), Army Research Office (W911NF-19-1-0032), and Air Force Office of Scientific Research (FA9550-16-1-0650).  

\section*{Declaration of interests}
\indent The authors report no conflict of interest.


\end{document}